\documentclass[11pt]{article}
\def\Vec#1{\mbox{\boldmath $ #1 $}}
\def\sVec#1{\mbox{\small\boldmath $ #1 $}}
\begin{document}
\begin{center}
\begin{large}
Electromagnetic Response Theory with Relativistic Corrections: \\
Selfconsistency and Validity of Variables  \\
\end{large}
Kikuo Cho (Sept 26, 2024) 
\end{center}

\underline{Abstract} \\
Schr\"odinger-Pauli equation (SP-eq) derived from weakly relativistic 
approximation (WRA) of Dirac eq, combined with Electromagnetic (EM) 
field Lagrangian for variational principle, is expected to give a new 
level of EM response theory.  A complete process of this formulation 
within the second order WRA is given, with explicit forms of charge and 
current densities, $\rho , \Vec{J}$, and electric and magnetic polarizations, 
$\Vec{P}$, $\Vec{M}$ containing correction terms.  They fulfill, not only
the continuity equation, but also the relations $\nabla \cdot \Vec{P}=-\rho, 
\ \partial \Vec{P}/\partial t + c \nabla \times \Vec{M} = \Vec{J}$, known 
in the classical EM theory for the corresponding macroscopic variables.  
This theory should be able to describe all the EM responses within the 
second order WRA, and the least necessary variables are ${\phi, \Vec{A}, 
\rho, \Vec{J}}$ (six independent components).  From this viewpoint, 
there emerges a problem about the use of "spin current" popularly  
discussed in spintronics, because it does not belong to the group of 
least necessary variables. \\

\section{Introduction}

EM response theory for atomic, molecular, and condensed matter physics 
needs Schr\"odinger and Maxwell equations with relativistic corrections. 
Among them, spin-orbit interaction often plays an essential role. 
In old days, its form is assigned to $\sim \Vec{\ell} \cdot \Vec{s}$, where 
$\Vec{\ell}$ and $\Vec{s}$ are orbital and spin angular momentum, respectively. 
Recently more details about it have been studied for the new 
class of problems about spintronics, multiferroics, etc. \cite{FroStu, 
Leurs, NagTok}, where relativistic corrections are studied in more detail.  

On the other hand, the aspect of selfconsistent (SC) motion of  interacting 
matter-EM field has been a main subject for resonant processes in 
exciton-polaritons and semiconductor nanostructures.  In this case 
the microscopic nonlocal character of EM response plays an essential role 
in the formulation, which also contributes to establish the hierarchy of 
EM response theories from QED to classical macroscopic one \cite{ChoNLRT}. 

As a fundamental EM response theory, both of \cite{FroStu, Leurs, NagTok}
and \cite{ChoNLRT} do not seem to be general enough. The former group 
does not pay much attention to the SC aspect, and the latter does 
not consider the relativistic corrections as much as the former does. 
The purpose of this note is to remedy the insufficiency by preparing the 
consistent set of quantum mechanical equation of electrons and M-eqs 
with relativistic correction terms.  For each order of WRA of Dirac eq, 
we prepare a set of SP-eq and M-eqs, to be handled as simultaneous equations.  
For specific problems the importance of the correction terms will be varying. 
Choosing appropriate part of the equations for each specific problem, we can 
use this scheme to handle a wide range of problems from a single point of view.  
 
We show the details of the SP-eq and M-eqs in the second order WRA 
below, confirming the continuity equation of charge and current densities. 
Also we demonstrate the decomposition of current density $\Vec{J}$ as 
$\partial \Vec{P}/\partial t + c\nabla \times \Vec{M}$ in terms of the 
explicitly defined operator forms of electric and magnetic polarizations,
$\Vec{P}$ and $\Vec{M}$, containing the correction terms of WRA. 
Since the decomposition of $\Vec{J}$ reflects the  
existence of orbital magnetic moment, or orbital angular momentum.  
its relevance to spin current will be discussed in the last section.

\section{Equations for EM response with relativistic corrections}

\subsection{Schr\"odinger-Pauli equation}
  
Dirac eq. for electron in EM field, can be written 
in the following form of two 2$\times$2 matrix equations 
for the two component wave functions $\psi_{-} = (\psi_{1}, \psi_{2})^{\rm T}$, 
$\psi_{+} = (\psi_{3}, \psi_{4})^{\rm T}$  $({\rm T=transposed})$ representing 
the negative and positive energy parts as 
\begin{eqnarray} 
\label{eq:Dirac3}
  (i\hbar\frac{\partial}{\partial t} - e\phi + 2 mc^2) \psi_{-} 
    &=& \Vec{\sigma}\cdot \Vec{\pi} \psi_{+}\ ,  \\
  (i\hbar\frac{\partial}{\partial t} - e\phi ) \psi_{+}
   &=& \Vec{\sigma}\cdot \Vec{\pi} \psi_{-} \ , 
\end{eqnarray}
where $e(<0), m, \Vec{p}$ are charge, mass, momentum of electron, respectively, 
$\phi, \Vec{A}$ scalar and vector potentials, $\Vec{\sigma}$ Pauli spin matrix, 
$\Vec{\pi} = c\Vec{p} - e\Vec{A}$, and the origin of energy is chosen at $+mc^2$. 
  
Eliminating $\psi_{-}$ from these equations, we have  
\begin{equation}
\label{eq:red-Dirac}
  \big[ (i\hbar\frac{\partial}{\partial t} - e\phi )  - \Vec{\sigma}\cdot\Vec{\pi}\ 
       ( i\hbar\frac{\partial}{\partial t} - e\phi + 2 mc^2)^{-1} 
         \Vec{\sigma}\cdot\Vec{\pi}) \big] \psi_{+} = 0  \ . 
\end{equation}
Weakly relativistic approximation (WRA) is the one where $mc^2$ is much larger 
than the energies of electron motion, EM field, and their interaction, which allows  
the power series expansion  
\begin{equation}
\label{eq:pow-exp}
  [i\hbar\frac{\partial}{\partial t} - e\phi + 2 mc^2]^{-1}  =  \frac{1}{2mc^2} 
        - \frac{1}{4m^2c^4} (i\hbar\frac{\partial}{\partial t} - e\phi) + \cdots  \ .
\end{equation}
Using this expansion up to a given order in eq.(\ref{eq:red-Dirac}), we obtain 
SP-eq, i.e.,  Schr\"odinger eq. with WRA-correction terms. 
The result of the first order approximation is 
\begin{eqnarray}
\label{eqn:1st-app}
 ( i\hbar \frac{\partial}{\partial t} - H_{P1})\ \psi_{+} = 0 \hspace{3cm}  &{}& \\
\label{eqn:Pauli1}
  H_{P1} = e\phi + \frac{1}{2m} (\Vec{p} - \frac{e}{c} \Vec{A})^2 
    - \frac{e\hbar}{2mc} \Vec{\sigma} \cdot \Vec{B}   \ ,  &{}&
\end{eqnarray}
The first order corrections are kinetic energy term and spin Zeeman term. 
The second order approximation leads to 
$i\hbar (\partial \psi_{+}/\partial t) - H_{P2} \psi_{+} = 0 $, where
\begin{equation}
\label{eqn:HP2}
  H_{P2}  =  \frac{1}{2mc^2} \Vec{\pi}^2 +  e\phi 
         - \frac{e\hbar}{2mc} \Vec{\sigma} \cdot \Vec{B}   
        - \frac{e\hbar}{8m^2c^3}\ [\Vec{\pi}\cdot(\Vec{\sigma} \times \Vec{E}) 
                  + (\Vec{\sigma} \times \Vec{E}) \cdot \Vec{\pi}]               
           - \frac{e\hbar^2}{8m^2c^2}\ \nabla\cdot\Vec{E}   \ .                          
\end{equation}
Second order corrections are spin-orbit interaction and Darwin term, i.e., 
the 4-th and the 5-th term on r.h.s. (Mass-velocity term appears 
in the third order correction.)  The form of the spin-orbit interaction 
given above is more complex than the traditional form  
$\sim \Vec{\ell} \cdot \Vec{s}$, because $\Vec{\pi}$ contains $\Vec{A}$ 
and $\Vec{E}$ contains its transverse component, in addition to the 
variables contributing to $\sim \Vec{\ell} \cdot \Vec{s}$.   

The existence of orbital Zeeman energy in $H_{P1}$ is known since early time 
of atomic spectroscopy \cite{Condon-S}.  It is shown from the following argument. 
The presence of static magnetic field $\Vec{B}$ can be described by 
$\Vec{A} = \Vec{B}\times\Vec{r}/2$.  
This allows to rewrite the $\Vec{A}$-linear term in the kinetic energy as 
\begin{equation}
-(e/2mc)(\Vec{p}\cdot\Vec{A}+ \Vec{A}\cdot\Vec{p}) = 
-(e/2mc) (\Vec{r}\times\Vec{p})\cdot\Vec{B}  \ , 
\end{equation}
which is orbital Zeeman energy. 

EM fields ($\Vec{E}, \Vec{B}$) or ($\Vec{A}, \phi$) are given quantities in Dirac 
eq. They can be of internal and/or external origin, which results in different 
physical problems.  For example, $\Vec{E}$ due to the core potential of atoms 
or ions leads to the traditional spin-orbit term in SP-eq. determining the energy 
levels of electrons in a matter system.  However, if it is an  applied external 
electric field, it will cause spin-Hall effect, i.e., up and down spins are swept 
to opposite directions along $\Vec{\sigma} \times \Vec{E}$.

\subsection{Maxwell equations under WRA}
\subsubsection{Charge and current densities}
The change in the Schr\"odinger equation due to WRA of Dirac eq. 
is reflected also to the EM field equations.  For its derivation under WRA, 
we can rely on the variation principle in terms of the Lagrangian density
\begin{equation}
\label{eqn:Meq}
  {\it L} = \psi^\dag (i\hbar \frac{\partial}{\partial t}  - H_{\rm P2}) \psi 
              + \frac{1}{8\pi} (E^2 - B^2) \ .
\end{equation}
The last term on r.h.s. is the contribution of free EM field. 
Hereafter $\psi_{+}$ is abbreviated as $\psi$.  

>From the condition that the action integral $\int\int{\rm d}\Vec{r}{\rm d}t {\it L}$ 
takes stationary value for the variations of $\phi$ and $\Vec{A}$, we obtain 
M-eqs in the familiar form
\begin{eqnarray}
\label{eqn:phimax}
&{}&  - \nabla \cdot \big[ \frac{1}{ c}  \frac{\partial \Vec{A}}{\partial t} 
            +  \nabla \phi \big] = 4\pi \rho \ , \\
\label{eqn:Amax}
&{}&  \frac{1}{c} \frac{\partial \nabla \phi}{\partial t} 
      + \frac{1}{c^2} \frac{\partial^2 \Vec{A}}{\partial t^2} 
      + \nabla \times \nabla \times\Vec{A}  = \frac{4\pi}{c} \Vec{J}  \ .
\end{eqnarray}
The charge and current densities are given, in the first order WRA, as 
\begin{eqnarray}
\label{eqn:chargeD1}
&{}&    \rho^{(1)} = \rho_{0}    \ ,      \\
\label{eqn:currentD1} 
&{}&  \Vec{J}^{(1)} = \Vec{J}_{0} + c \nabla \times \Vec{M}_{spin}  \ , 
\end{eqnarray}
where 
\begin{equation} 
  \rho_{0} =  e \psi^{\dagger} \psi\ , \ \ \ 
  \Vec{J}_{0} =  \frac{e}{mc} \psi^{\dagger} \Vec{\pi} \psi \ , \ \ \ 
  \Vec{M}_{spin} = \frac{e\hbar}{2mc} \psi^{\dagger} \Vec{\sigma} \psi \ .
\end{equation}
In the second order WRA, they are 
\begin{eqnarray}
\label{eqn:chargeD}
&{}&    \rho^{(2)} = \rho_{0}  - \nabla \cdot (\Vec{P}_{SO} + \Vec{P}_{D})  \ , \\
\label{eqn:currentD}                                                                                         
&{}&  \Vec{J}^{(2)} =  \Vec{J}_{0} - e \Vec{F} + c \nabla \times \Vec{M}_{spin}   
                  + \frac{\partial}{\partial t}(\Vec{P}_{SO} + \Vec{P}_{D}) \ , 
\end{eqnarray}
where
\begin{equation}
           \Vec{F} =  \frac{1}{2mc} \Vec{M}_{spin} \times \Vec{E} \ , \ \ \   
           \Vec{P}_{SO} = \frac{e\hbar}{4m^2c^3} 
                 \psi^{\dagger} (\Vec{\pi} \times \Vec{\sigma}) \psi   \ , \ \ \ 
           \Vec{P}_{D} = - \frac{e\hbar^2}{8m^2c^2}\ \nabla \psi^{\dagger} \psi \ .  
\end{equation}
It should be noted that the charge density is, not only that of the electrons 
and nuclei in a sample, but also possibly of external charges.

\subsubsection{Continuity equation}
It is well known that the continuity equation holds 
between charge and current densities 
in both limits of classical and fully relativistic cases.  
It may be of interest to see if it is 
also valid under WRA.  

>From the result of the previous section, we get 
\begin{eqnarray}
  \frac{\partial \rho^{(1)}}{\partial t} + \nabla \cdot \Vec{J}^{(1)} 
     &=& \frac{\partial \rho_{0}}{\partial t} + \nabla \cdot \Vec{J}_{0}  \ , \\
  \frac{\partial \rho^{(2)}}{\partial t} + \nabla \cdot \Vec{J}^{(2)} 
     &=& \frac{\partial \rho_{0}}{\partial t} 
            + \nabla \cdot (\Vec{J}_{0} - e \Vec{F}) \ .
\end{eqnarray}
Time evolution $\partial \rho_{0}/\partial t$  in the case of second order WRA 
is calculated via SP eq with $H_{P2}$ as  
\begin{eqnarray}
  \frac{\partial }{\partial t} e \psi^\dagger \psi \ \
         &=& e \big[\frac{\partial \psi^\dagger }{\partial t} \psi 
         +   \psi^\dagger \frac{\partial \psi}{\partial t} \big]  
          = \frac{ie}{\hbar} \big[(H_{P2}\psi)^\dagger \psi -  \psi^\dagger (H_{P2} \psi) \big]   \\
    &=& \frac{ie\hbar}{2m} ( \psi^\dagger \nabla^2 \psi  
                                      - \psi \nabla^2 \psi^\dagger )   
          +  \frac{e^2}{2mc}  \psi^\dagger (\nabla \cdot \Vec{A} 
                                             + \Vec{A} \cdot \nabla) \psi  \nonumber  \\ 
    &{}&    - \frac{e^2 \hbar}{8m^2c^2}\psi^\dagger  \{(\nabla \times \Vec{\sigma}) \cdot \Vec{E} 
                               + \Vec{E} \cdot (\nabla \times \Vec{\sigma}) \}\psi  \nonumber \\
      &=& \nabla \cdot \big[\frac{ie\hbar}{2m} 
                                (\psi^\dagger \nabla \psi - \psi \nabla \psi^\dagger )  
       + \frac{e^2}{mc} \psi^\dagger \Vec{A} \psi   - \frac{e^2\hbar}{4m^2c^3} 
                        \psi^\dagger (\Vec{\sigma} \times \Vec{E})\psi \big]  \\
      &=& - \nabla \cdot (\Vec{J}_{0}  - e \Vec{F}) \ .                           
\end{eqnarray}  
This is the continuity equation for the 2nd order WRA. which includes the  
case of  the first order WRA ($\Vec{F}=0$).

\subsubsection{Coulomb Potential}
 
Equation (\ref{eqn:phimax}) is the Gauss law $\nabla \cdot \Vec{E} = 4\pi \rho$ 
with corrected charge density.   This means that the Coulomb potential between 
charge densities also contains the contribution of WRA. The solution of this equation 
gives the longitudinal (L) part of $\Vec{E}$ 
\begin{equation}
  \Vec{E}_{\rm L}(\Vec{r}) = - \nabla \int{\rm d}\Vec{r}' 
                        \frac{\rho(\Vec{r}')}{|\Vec{r}-\Vec{r}'|} \ . 
\end{equation}
In Coulomb gauge, we have $\nabla \cdot \Vec{A} = 0$ and $\Vec{E}_{L} = -\nabla \phi$.  
The selfenergy of the longitudinal field can be rewritten as 
\begin{equation}
\label{eqn:L-en}
  \frac{1}{8\pi} \int E_{L}^2 {\rm d}\Vec{r} 
    = \frac{1}{2} \int \int {\rm d}\Vec{r} {\rm d}\Vec{r}' 
            \frac{\rho(\Vec{r})\rho(\Vec{r'}) }{|\Vec{r}-\Vec{r}'|}
    = \frac{1}{2} \int{\rm d}\Vec{r} \rho \phi   \ .
\end{equation}
On the other hand, $\phi$ dependent terms in $H_{P2}$ are 
\begin{equation}
\label{eqn:HP2phi}
   \rho_{0}\phi  - \frac{e\hbar}{4m^2c^3}\ 
               \phi \nabla \cdot \psi^{\dagger}  (\Vec{\pi}\times\Vec{\sigma})\psi                             
                + \frac{e\hbar^2}{8m^2c^2}\psi^{\dagger} 
                 \nabla\cdot\nabla \phi \psi = \rho \phi \ .                       
\end{equation}
In the action integral, half of it cancels the selfenergy eq(\ref{eqn:L-en}), resulting 
in the Coulomb potential with the charge density containing WRA corrections.  
Without the correction terms, it gives the classical Coulomb potential.

\section{Electric and Magnetic Polarizations}

In classical electromagnetics, the electric and magnetic polarizations are known to 
satisfy $\nabla\cdot \Vec{P} = -\rho$, $\Vec{J} = (\partial \Vec{P}/\partial t) + 
c \nabla \times \Vec{M}$, and this is used to rewrite Amp\`ere law from microscopic 
to macroscopic forms.  Thereby macroscopic variables $\Vec{D} = \Vec{E} + 
4\pi \Vec{P}$ and  $\Vec{H} = \Vec{B} - 4\pi \Vec{M}$ are introduced instead of 
$\Vec{E}$ and $\Vec{B}$.  Traditional arguments 
to derive these relations does not seem to be general and rigorous enough. 
It is mainly due to the ambiguously defined component of current density, 
such as conduction current density,   
which cannot be classified into $\Vec{P}$ and $\Vec{M}$. 
 A clear cut definition of $\Vec{P}$ and $\Vec{M}$ satisfying the relations 
is given by Cohen-Tannoudji et al (CT)  \cite{Cohen-T}.  For a charged particle 
system with charge neutrality as a whole, where $\Vec{r}_{\ell}, \Vec{v}_{\ell}$ are 
the coordinate and velocity of $\ell$-th particle, they showed that
\begin{eqnarray}
\label{eqn:CTCha}
  \rho(\Vec{r}) &=& \sum_{\ell} e_{\ell} \delta(\Vec{r} - \Vec{r}_{\ell}) \ , \\
\label{eqn:CTCur}
  \Vec{J}(\Vec{r}) &=& \sum_{\ell} e_{\ell} \Vec{v}_{\ell} 
                                   \delta(\Vec{r} - \Vec{r}_{\ell})           \ , \\
\label{eqn:CTP}
\Vec{P}(\Vec{r}) &=& \int_{0}^{1} du \sum_{\ell} e_{\ell} \Vec{r}_{\ell} 
                                   \delta(\Vec{r} - u \Vec{r}_{\ell})         \ , \\  
\label{eqn:CTM}
\Vec{M}(\Vec{r}) &=& \frac{1}{c} \int_{0}^{1} udu \sum_{\ell} 
                                e_{\ell} \Vec{r}_{\ell} \times \Vec{v}_{\ell}
                                   \delta(\Vec{r} - u \Vec{r}_{\ell})
\end{eqnarray} 
satisfy $\nabla \cdot \Vec{P}= -\rho$ and $\Vec{J} = (\partial \Vec{P}/\partial t) 
+ c \nabla \times \Vec{M}$.  \\

Now, we ask whether similar relations can be shown in the presence of relativistic 
corrections.  We assume a charge neutral system as a whole, following the case of CT. \\

\underline{First order WRA}\\

The charge density in the first order WRA is $\rho_{0} = e\psi^{\dagger}\psi$. 
The solution of $\nabla \cdot \Vec{P}_{0} = - \rho_{0}$ in 1D space would be 
$-\int^{x} \rho_{0} d\bar{x} $, which, however, cannot be used in 3D case.
By the help of (\ref{eqn:CTP}), let us assume  
\begin{equation}
\label{eqn:ChoP}
  \Vec{P}_{0}(\Vec{r}) =  \int_{0}^{1}{\rm d}u \int{\rm d}\bar{\Vec{s}} \ 
                 \rho_{0}(\bar{\Vec{s}})\ \bar{\Vec{s}} \ \delta(\Vec{r} - u\bar{\Vec{s}}) \ , 
\end{equation}
where $\bar{\Vec{s}}$ is a coordinate variable for additional integral, and 
$\Vec{v}_{s}$, which appears later, is its corresponding velocity. 
(For $\Vec{r}$, velocity is $\Vec{\pi}/mc$.) 
The $\Vec{k}$-th Fourier component of  $\nabla \cdot \Vec{P}_{0}$ is 
\begin{eqnarray}
 i\Vec{k}\cdot\Vec{P}_{0\sVec{k}} &=& \int_{0}^{1}{\rm d}u \int{\rm d}\bar{\Vec{s}} 
        \  \rho_{0}\ (i\Vec{k}\cdot\bar{\Vec{s}})  
  \exp(-i\Vec{k}\cdot\bar{\Vec{s}} u)      \nonumber     \\
   &=& - \int{\rm d}\bar{\Vec{s}}\ \ \rho_{0} \  
               \exp(-i\Vec{k}\cdot\bar{\Vec{s}}) \ , 
\end{eqnarray}
where we consider charge neutrality $\int d\bar{s}\ \rho_{0} = 0$.  This result is 
$\nabla \cdot \Vec{P}_{0} = -\rho_{0}$ in the coordinate space, demonstrating 
the validity of the operator form (\ref{eqn:ChoP}). 

Spin Zeeman term of the current density in the first order WRA is 
already written as "rotation of spin magnetization".  Thus, the remaining problem  
is to find $\Vec{M}^{(1)}$ satisfying 
 $\Vec{J}_{0} = (\partial \Vec{P}_{0}/\partial t) + c \nabla \times \Vec{M}^{(1)}$, 
which corresponds to the case of classical particle system. 
 
Time evolution of $\Vec{P}_{0\sVec{k}}$ due to $H_{P1}$ is 
\begin{equation}
 [\partial \Vec{P}_{0\sVec{k}}/\partial t]_{1}
            = - e\int_{0}^{1}du \int{\rm d}\bar{\Vec{s}}
              (\nabla \cdot \psi^{\dagger} \Vec{v}_{s} \psi) \bar{\Vec{s}} \ 
              \exp(-i\Vec{k}\cdot\bar{\Vec{s}} u)   \ .
\end{equation}
This can be rewritten, via partial integration, as 
\begin{eqnarray}
      [\partial \Vec{P}_{0\sVec{k}}/\partial t]_{1} = 
       e \int_{0}^{1}du \int{\rm d}\bar{\Vec{s}}  \left[\psi^{\dagger} \Vec{v}_{s} \psi 
                         - i\Vec{k}\cdot \psi^{\dagger} \Vec{v}_{s} \psi  \bar{\Vec{s}} \right] 
                          \  \exp(-i\Vec{k}\cdot\bar{\Vec{s}} u) \ . 
\end{eqnarray}
The contribution of the primitive function  
 $[\psi^{\dagger} \Vec{v}_{s} \psi) \bar{\Vec{s}} \ 
              \exp(-i\Vec{k}\cdot\bar{\Vec{s}} u)] $
at the lower and upper limits of integration range vanishes by taking the range  
so large that matter field has zero amplitude at the limits.  
Following eq.(29), we may assume $\Vec{M}^{(1)}$ in the form
\begin{equation}
\label{eqn:Mpol}
  \Vec{M}^{(1)}(\Vec{r}) = \frac{e}{c}\int_{0}^{1} u du \int{\rm d} \bar{\Vec{s}}\    
    \psi^{\dagger}(\bar{\Vec{s}}) (\bar{\Vec{s}} \times \Vec{v}_{s})  \psi(\bar{\Vec{s}})       
     \delta(\Vec{r} - u\bar{\Vec{s}}) \ , 
\end{equation}
which leads to 
\begin{equation}
     (c \nabla \times \Vec{M}^{(1)})_{\sVec{k}}
      = ie \int_{0}^{1} u{\rm d}u \int{\rm d}\bar{\Vec{s}}   
    \ \psi(\bar{\Vec{s}})^{\dagger} (\Vec{k} \times\bar{\Vec{s}} \times \Vec{v}_{s}) 
                   \psi(\bar{\Vec{s}}) \ \exp(-i\Vec{k}\cdot\bar{\Vec{s}} u)    \nonumber \\  
\end{equation}
The vector triple product is decomposed into $\Vec{k} \times\bar{\Vec{s}} \times \Vec{v}_{s} 
= (\Vec{k}\cdot\Vec{v}_{s}) \bar{\Vec{s}} - (\Vec{k}\cdot\bar{\Vec{s}}) \Vec{v}_{s}$, where 
the contribution of $(\Vec{k}\cdot\Vec{v}_{s}) \bar{\Vec{s}} $ cancels the second term on 
the r.h.s. of $[\partial \Vec{P}_{0\sVec{k}}/\partial t]_{1}$, 
and $(\Vec{k}\cdot\bar{\Vec{s}}) \Vec{v}_{s}$ gives 
\begin{equation}
     e \int{\rm d}\bar{\Vec{s}} \ \psi(\bar{\Vec{s}})^{\dagger}  \Vec{v}_{s} \psi(\bar{\Vec{s}}) \
               \int_{0}^{1} u{\rm d}u \frac{d}{du} \exp(-i\Vec{k}\cdot\bar{\Vec{s}} u) \ . 
\end{equation}
The $u$ integration gives 
\begin{equation}
     \int_{0}^{1} u{\rm d}u \frac{d}{du} \exp(-i\Vec{k}\cdot\bar{\Vec{s}} u) 
                = \exp(-i\Vec{k}\cdot\bar{\Vec{s}})                           
    -  \int_{0}^{1} {\rm d}u \exp(-i\Vec{k}\cdot\bar{\Vec{s}} u) \ .
\end{equation} 
The contribution of the second term on the r.h.s. cancels the first term of 
$[\partial \Vec{P_{0\sVec{k}}}/\partial t]_{1}$.  
What remains finally is the desired result 
\begin{equation}
  [\partial \Vec{P}_{0\sVec{k}}/\partial t]_{1}
                    + (c \nabla \times \Vec{M}^{(1)})_{\sVec{k}}
            = e \int{\rm d}\bar{\Vec{s}} \nonumber \\
  \ \psi(\bar{\Vec{s}})^{\dagger} \Vec{v}_{s} \psi(\bar{\Vec{s}}) 
      \ \exp(-i\Vec{k}\cdot\bar{\Vec{s}}) 
       = [\Vec{J}^{(1)}_{\sVec{k}}] \ .
\end{equation} 

\underline{Second order WRA}\\

The electric polarization due to $\rho_{0}$ is $\Vec{P}_{0}$ as in the first 
order case, but its time evolution is driven, not by $H_{P1}$, but by $H_{P2}$, 
which gives 
\begin{equation}
 [\partial \Vec{P}_{0\sVec{k}}/\partial t]_{2}
            = e\int_{0}^{1}du \int{\rm d}\bar{\Vec{s}}
       [\nabla \cdot \psi^{\dagger} (\Vec{v}_{s} - \Vec{F}) \psi] \bar{\Vec{s}} \ 
        \exp(-i\Vec{k}\cdot\bar{\Vec{s}} u)   \ .
\end{equation}
This result is same as $ [\partial \Vec{P}_{0\sVec{k}}/\partial t]_{1}$ except for  
replacing $\Vec{v}_{s}$ by $\Vec{v}_{s} - \Vec{F}$. Same replacement occurs 
between $\Vec{J}^{(1)}$ and  $\Vec{J}^{(2)}$.  
This suggests us to define the second order $\Vec{M}^{(2)}$ by the same replacement 
in $\Vec{M}^{(1)}$ as 
\begin{equation}
\label{eqn:Mpol2}
  \Vec{M}^{(2)}(\Vec{r}) = \frac{e}{c} \int_{0}^{1} u du \int{\rm d} \bar{\Vec{s}}  
    \psi^{\dagger}  [\bar{\Vec{s}} \times (\Vec{v}_{s} - \Vec{F})]  \psi\ 
      \delta(\Vec{r} - u\bar{\Vec{s}}) \ .
\end{equation}
Then, the argument for the first order can be applied to the second order 
resulting in  
\begin{equation}
  \Vec{J}_{0} - e\Vec{F} =  [\partial \Vec{P}_{0\sVec{k}}/\partial t]_{2} 
                    + c \nabla \times \Vec{M}^{(2)} \ . 
\end{equation}
Adding $\partial(\Vec{P}_{SO} + \Vec{P}_{D})/\partial t + 
c \nabla \times \Vec{M}_{spin}$ 
to both hand sides, we have $\Vec{J}^{(2)}$ on the l.h.s.  
In this way, we establish the relation 
 $\Vec{J} = (\partial \Vec{P}/\partial t) + c \nabla \times \Vec{M}$ 
in each order of WRA. \\
 
\section{Discussions}

 \subsection{General Framework of EM response theory under WRA}
 
The present formalism is the one to provide the fundamental equations of  EM 
response of matter with consideration of all the possible correction terms of 
the second order WRA of Dirac eq.  It consists of SP-eq,  
$i\hbar (\partial \psi/\partial t) - H_{P2} \psi = 0 $,  
and M-eqs,  (\ref{eqn:phimax}) and (\ref{eqn:Amax}),  with $\rho, \Vec{J}$ of 
(\ref{eqn:chargeD}) and (\ref{eqn:currentD}). 
The variables for the description are minimal necessary ones, 
$\{\rho, \Vec{J}, \phi, \Vec{A}\}$, consisting of six independent components.  
>From SP-eq and initial condition of matter we can calculate the expectation 
values of $\{\rho, \Vec{J}\}$, which play the role of the source terms of M-eqs.  
Solving the SP-eq and M-eqs simultaneously, one determines the six variables,   
as functions of time and position.  Since there is no other variable in the two 
fundamental equations, this should contain all the necessary information 
about the measurable quantities of EM response. 
This is a general feature common to all the EM response theories from QED to 
classical one, and it will lead to a conflict with the use of additional variable 
"spin current" as a measurable physical quantity of EM response.  This will be 
discussed in the next subsection.    

It is possible to choose alternative variables $\{\Vec{P}, \Vec{M}\}$ instead of 
$\{\rho, \Vec{J}\}$, based on the operator equations $\nabla \cdot \Vec{P} = -\rho$
and $\Vec{J}  = \partial \Vec{P}/\partial t + c\nabla \times \Vec{M}$.  These relations, 
known in the classical EM theory, is revisited in the presence of the correction terms 
of WRA in Dirac- and M-eqs, and are rigorously confirmed as operator equations. 
The forms of $\Vec{P}$ and $\Vec{M}$ are exactly given.  They are, in the lowest order, 
standard electric polarization and \{orbital and spin\} magnetization, but the second 
order corrections bring further mixing of orbital and spin. It is nevertheless remarkable 
that the mathematical expressions of $\Vec{P}$ and $\Vec{M}$ satisfying $\Vec{J}  = 
\partial \Vec{P}/\partial t + c\nabla \times \Vec{M}$ can be found. \\

In order to allow wide range of applicability, we keep all the (second order) correction 
terms of WRA in SP- and M-eqs.  On applying to a specific problem, the importance 
of each correction term will be varying, so that the selection of important ones will be 
useful to avoid unnecessary complication.  Among many possible cases, we mention 
here one dividing point about the selection.  This is whether the SC solution is 
required or not.  In the case of NLRT \cite{ChoNLRT} dealing with the 
polaritons and nano-scale optical responses, SC solution plays an essential 
role to describe the resonant behavior, which may lead to remarkable  
dependence on sample size and shape. This is a typical behavior of 
self-sustaining modes of resonant states. 

On the other hand, the problems treated in modern spin related phenomena are 
described by SP-eq with various WRA corrections \cite{FroStu, Leurs, NagTok}, 
where little is mentioned about the corresponding M-eqs. An exception is 
\cite{Wang}, who derived the correction terms of charge density via variation 
principle, but not the orbital magnetization (or angular momentum) in contrast  
to our result in Sec.2, 3.  Correspondingly, no intension is seen to treat SP-eq 
and M-eqs as simultaneous equations for EM response, neither to restrict the 
variables to the least necessary one.

\subsection{Spin density and spin current ?} 

The spin density and spin current may be compared with charge density $\rho$ 
and (charge) current density $\Vec{J}$.  The definition of the latter is very clear 
in any order of WRA, i.e., the variation of action integral of $L$ with respect 
to $\phi$ and $\Vec{A}$ gives $\rho$ and $\Vec{J}$, respectively. A similar 
definition for spin density and spin current for the same $L$ does not work well.  
To get the spin density $\psi^{\dagger} \Vec{\sigma} \psi$ in the first order WRA, 
one will take the variation with respect to $\Vec{B}$, but this brings about 
orbital angular momentum additionally.  No variational principle seems to 
exist for spin and orbital angular momenta separately.  The mixing of spin and 
orbital angular momenta occurs further in the second order WRA.  

To avoid the difficulty of spin current,  one might think of a simplified model 
of electronic states consisting of only s-electrons ($\ell =0$).  
In order to define spin density and its time derivative (spin current) 
via variational principle, we need "matter Hamiltonian of s-orbital alone".  
Obviously, such an operator does not exist, so that even this 
simplified model does not support the parallelism with $\rho$ and $\Vec{J}$.  
 
 As we mentioned in the previous subsection, we have six independent equations 
for six independent variables $\{\rho, \Vec{J}, \phi, \Vec{A}\}$ 
as the fundamental equations for EM response.  
The solutions of these simultaneous equations should describe all the possible 
situations of EM response.  Since this statement is applicable to any 
specific problem of EM response, no additional variables are required.   
In the field of spintronics, "spin current" seems to be a principal variable. 
It is independent from $\rho, \Vec{J}$, so that it is not compatible with 
the statement given above.   There are some works proposing  different definitions 
of spin current \cite{Wang, An}, which suggests insufficient establishment of 
the concept, or its illegitimacy in EM response theory.   From the viewpoint 
of the present scheme, there should be a way to describe spin Hall effect without 
referring to spin current, which might be worth trying. \\

Comments by Mr. S. Inoue and Dr. T. Mii are acknowledged as to the typing 
errors and signs of some formulas in the firstl manuscript.  \\

\end{document}